\documentclass[	DIV=calc,%
							paper=letter,%
							fontsize=11pt,%
							twocolumn]{scrartcl}	 					

\usepackage{lipsum}
\usepackage{wrapfig}
\usepackage{hyperref}										
\usepackage[english]{babel}										
\usepackage[top=2.7cm, bottom=2.4cm, left=2.15cm, right=2.15cm]{geometry}
\usepackage[protrusion=true,expansion=true]{microtype}				
\usepackage{amsmath,amsfonts,amsthm}					
\usepackage[pdftex]{graphicx}									
\usepackage[svgnames]{xcolor}									
\usepackage[hang, small,labelfont=bf,up,textfont=it,up]{caption}	
\usepackage{epstopdf}												
\usepackage{subfig}													
\usepackage{booktabs}												
\usepackage{fix-cm}													

\usepackage{sectsty}													
\allsectionsfont{
	\usefont{OT1}{phv}{b}{n}
	}

\sectionfont{
	\usefont{OT1}{phv}{b}{n}
	}

\usepackage{fancyhdr}												
\usepackage{lastpage}	

\usepackage{lettrine}
\newcommand{\initial}[1]{%
     \lettrine[lines=3,lhang=0.3,nindent=0em]{
     				\color{DarkGoldenrod}
     				{\textsf{#1}}}{}}

\usepackage{titling}															

\newcommand{\HorRule}{\color{DarkGoldenrod}
									  	\rule{\linewidth}{1pt}\vskip 0em%
										}

\pretitle{\vspace{-45pt} \begin{flushleft} \HorRule
				\fontsize{20}{20} \usefont{OT1}{phv}{b}{n} \color{DarkRed} \selectfont
				}
\title{Individualism: The legacy of great physicists}					
\posttitle{\par\end{flushleft}\vskip 0em}
\preauthor{\begin{flushleft}
					\large \lineskip 0em \usefont{OT1}{phv}{b}{sl} \color{DarkRed}}
\author{Ricardo Heras }											
\postauthor{\footnotesize \usefont{OT1}{phv}{m}{sl} \color{Black}
					 	\\\textbf{Department of Physics and Astronomy, University College London}
                         \\Published online in Physics Today, Points of View, October 2013 \href{http://scitation.aip.org/content/aip/magazine/physicstoday/news/10.1063/pt.5.2003}{(\textcolor[rgb]{0.00,0.00,1.00}{URL})}
                         \\ \footnotesize  \textcolor[rgb]{0.00,0.00,1.00}{ricardo.heras.13@ucl.ac.uk}
\par\end{flushleft}\vskip -1em	\HorRule}

\date{}																				

\begin{document}
\maketitle
\thispagestyle{fancy} 			
\initial{M}{ax Planck, the father of quantum physics, once said:\emph{``New scientific ideas never spring from a communal body, however organized, but rather from the head of an individually inspired researcher who struggles with his problems in lonely thought and unites all his thought on one single point which is his whole world for the moment." }\vspace{3 mm}
\\
Like Planck I believe that physics students should pursue individualism and independent thought, since those traits promote qualities needed for original research, such as a natural curiosity, reasonable doubt, a passion for understanding nature, and most importantly, an imaginative and creative mind.
\vspace{3 mm}
\\
However, advisers and mentors have stressed that collaboration is valuable for research, and is the better method to becoming a good scientist. I also have often heard that collaboration promotes professional connections and develops job opportunities, both of which are important for young scientists. So, should individualism or collectivism be encouraged in physics students?
\vspace{3 mm}
\\
Individualism stresses the importance and worth of independent thought, and encourages us to focus on our own goals and aspirations. The individualistic thinker pursues his or her own desires without interference from others. In the case of scientists, individualism means achieving personal gratification through understanding nature. That goal sows in them the seeds of the scientific passion necessary for a life-long commitment to research. The passionate scientist is someone who believes that the study, reflection, and understanding of nature are joys in and of themselves.
\vspace{3 mm}
\\
For such a scientist, rewards received in the course of his or her career--such as scientific prestige or financial benefits--cannot match the great intellectual pleasure of understanding the natural world. A scientist may be compared to a musician, who strives to harmonize each melody, tempo, and instrument in order to express his or her passion through a beautiful sonata, symphony, or other piece.
\vspace{3 mm}
\\
I am convinced that independent thought awakens creativity and imagination in physics students. Indeed, the very process of devotion to solitary rational thought seems to spark the related processes of imagination and creativity. Our minds seek consistency between previously established ideas, the development of possible solutions, and predictions emerging from these solutions. When we find that consistency, our minds work to create an imaginative solution for any proposed problem. As the mathematician Henri Poincar\'e claimed: \emph{``The mind uses its faculty for creativity only when experience forces it to do so."}
\vspace{3 mm}
\\
The development of independent thought stimulates students to value their own observation, experience, and intuition. They learn to be curious and to challenge their own assumptions. Scientists ought to seek truth, and in order to do so, they must learn to doubt.
\vspace{3 mm}
\\
Curiosity and doubt are two legacies of great, individually-inspired physicists. Albert Einstein said: \emph{``The important thing is not to stop questioning; curiosity has its own reason for existing...The important thing is not to stop questioning; never lose a holy curiosity." } If Isaac Newton had not been curious about why things fall, he would not have developed his universal law of gravitation. Likewise, if Einstein had not asked what a beam of light would look like if he caught up with it, he would not have developed his theory of special relativity.
\begin{figure*}[htb]
\centering
\includegraphics[width=.80\textwidth]{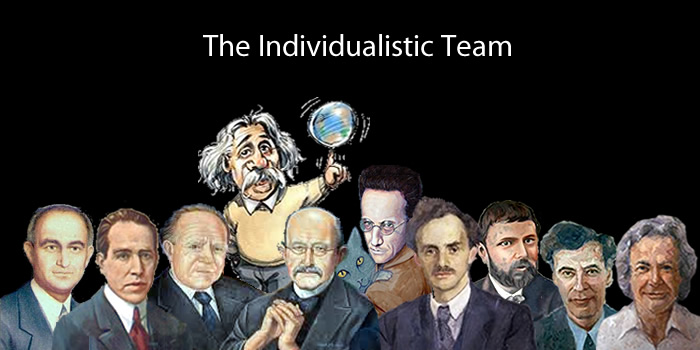}\\
\emph{\small\sffamily \textcolor[rgb]{0.24,0.24,0.24}{Notable individualistic physicists. From left to right: Enrico Fermi, Niels Bohr, Werner \quad\quad\quad\quad\quad\\ Heisenberg, Albert Einstein, Max Planck, Erwin Schr\"odinger, Paul Dirac, Henri Poincar\'e,\quad\quad\quad\quad \\Lev Landau, and Richard Feynman.}\quad\quad\quad\quad\quad\quad\quad\quad\quad\quad\quad\quad\quad\quad\quad\quad\quad\quad\quad\quad\quad\quad\quad\quad}
\end{figure*}
\vspace{3 mm}
\\
But individualism in science seems to be disappearing. Recent studies suggest that single authors seem to be an endangered species. In his Nature essay, \emph{``The demise of the lone author,"} Mott Greene observed [\href{http://www.nature.com/nature/journal/v450/n7173/full/4501165a.html}{\textcolor[rgb]{0.00,0.00,1.00}{1}}]: \emph{``Any issue of Nature today has nearly the same number of Articles and Letters as one from 1950, but about four times as many authors. The lone author has all but disappeared."} In his short arXiv communication, \emph{``Publication trends in astronomy: The lone author,"} Edwin A. Henneken [\href{http://arxiv.org/abs/1202.4646}{\textcolor[rgb]{0.00,0.00,1.00}{2}}] concluded that astronomers have dramatically trended away from single-author papers. And in his Physics Today commentary \emph{``Too many authors, too few creators,"} Philip J. Wyatt [\href{http://scitation.aip.org/content/aip/magazine/physicstoday/article/65/4/10.1063/PT.3.1499}{\textcolor[rgb]{0.00,0.00,1.00}{3}}] argued that individual creative researchers have declined since the second half of the 20th century. He wrote,  \emph{``The results seem truly astonishing. Although the data sets selected are relatively small, they show the downward trend of individual creativity. Most of the papers studied were written by authorship, or associated with, academia."}
\vspace{3 mm}
\\
Wyatt's worrying conclusion suggests that collective papers have increased at the expense of individual creativity. Funding agencies and institutions have motivated the proliferation of multiple-author papers. But in his reply to Greene's Nature essay, Kevin Hallock [\href{http://www.nature.com/nature/journal/v452/n7185/full/452282e.html}{\textcolor[rgb]{0.00,0.00,1.00}{4}}] pointed out that qualities of a lone author are beneficial to science. \emph{``I believe that funding agencies and institutions should also encourage single-author papers,"} he wrote. \emph{``The effort and initiative required to publish alone suggests an independent and tenacious scientist--both highly desirable qualities in any researcher."}
\vspace{3 mm}
\\
I do not mean to reject collective research, which is also crucial to natural sciences. Some of the large, experimental setups, at CERN or Fermilab, for example, require complete collaboration between hundreds of physicists and could not be carried through in any other way. Rather, I mean to suggest a renewed commitment to an essential requirement for creative research.
\vspace{3 mm}
\\
The best collective research emerges when independent scientists come together. Francis Crick and James Watson's quest for the DNA structure illustrates very well how collaboration and individual creativity merge on the path to scientific achievement. As Crick recounted: \emph{``Both of us had decided, quite independently of each other, that the central problem in molecular biology was the chemical structure of the gene."}
\vspace{3 mm}
\\
I disagree with the idea of a formal education in which students are not free to explore their own ideas and passions. In that model, students become too dependent upon their teachers and advisers, and curiosity and imagination have little space. I believe this is what Einstein was referring to when he said, \emph{``It is a miracle that curiosity survives formal education."}
\vspace{3 mm}
\\
Good science professors not only encourage their pupils to think for themselves, but they also simultaneously foster their students' joy in understanding the physical world. Professors and students should nurture relationships where creative learning is a priority. As Einstein said: \emph{``It is the supreme art of the teacher to awaken joy in creative expression and knowledge."}
\vspace{80 mm}
\\
In his Physics Today commentary, Wyatt reflected on an educational practice that I feel that most advisers should know and adopt:\emph{ ``Enrico Fermi, wanting to encourage individual creativity and innovation, required his PhD students to select their problem, solve it, and submit the results for publication in their name alone."}
\begin{figure*}[htb]
\centering
\includegraphics[width=.77\textwidth]{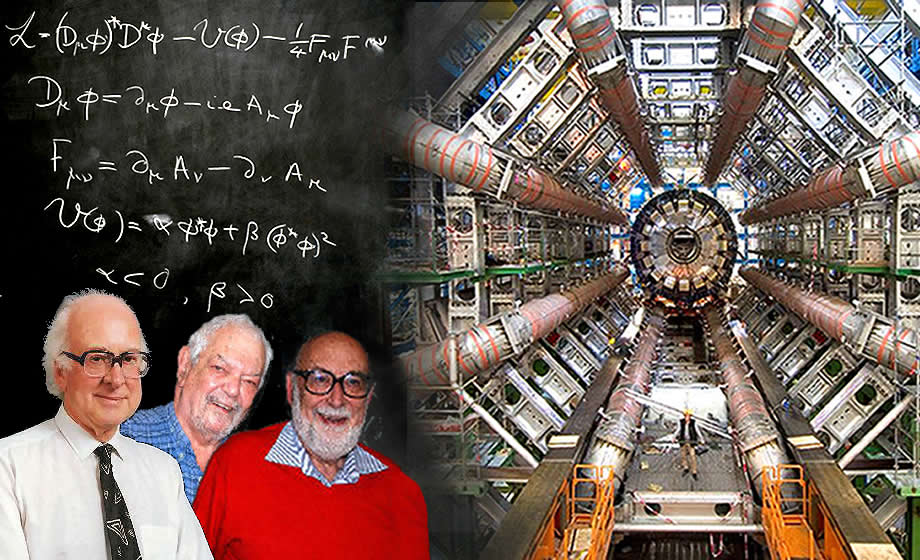}\\
\emph{\small \sffamily \textcolor[rgb]{0.24,0.24,0.24}{ A huge experimental collaboration of thousands of scientists at the Large Hadron Collider \quad\quad \\was needed last year to find the Higgs boson. But the idea behind this massive particle\quad\quad\quad \\ was the result of a handful of individuals, among them (left to right) Peter Higgs, Robert \\Brout, and Fran\c{c}ois Englert. Clearly, theoretical and applied physics are necessary and \quad\quad\quad\quad\\complementary. Although, as Englert once said: ``If you only do applied research, you \quad\quad\quad\quad  \\quickly lose creativity." } \quad\quad\quad\quad\quad\quad\quad\quad\quad\quad\quad\quad\quad\quad\quad\quad\quad\quad\quad\quad\quad\quad\quad\quad\quad\quad\quad\quad}
\end{figure*}{}

\end{document}